\documentclass[structabstract]{aa}

\usepackage{amsfonts}
\usepackage{times}
\usepackage{xspace,amssymb,amsmath,tabularx}

\usepackage{color}
\usepackage[bookmarksopen=false,bookmarksnumbered=true,breaklinks=true,
            colorlinks=true,linkcolor=blue,citecolor=blue]{hyperref}

\usepackage{graphicx}

\DeclareGraphicsExtensions{.eps}
\graphicspath{{./figures/}}
\usepackage{natbib}
\usepackage{array}
\usepackage{float}

\newcommand*{\bphi}{\boldsymbol{\phi}}
\newcommand*{\balpha}{\boldsymbol{\alpha}}
\newcommand*{\bbeta}{\boldsymbol{\beta}}

\newcommand*{\hbphi}{\hat{\boldsymbol{\phi}}}
\newcommand*{\bi}{\boldsymbol{i}}
\newcommand*{\bh}{\boldsymbol{h}}
\newcommand*{\bs}{\boldsymbol{s}}
\newcommand*{\bn}{\boldsymbol{n}}
\newcommand*{\bv}{\boldsymbol{v}}
\newcommand*{\bsigma}{\boldsymbol{\sigma}}
\newcommand*{\bm}{\boldsymbol{m}}
\newcommand*{\bT}{\boldsymbol{T}}
\newcommand*{\bD}{\boldsymbol{D}}
\newcommand*{\bC}{\boldsymbol{C}}
\newcommand*{\bM}{\boldsymbol{M}}
\newcommand*{\bN}{\boldsymbol{N}}
\newcommand*{\bF}{\boldsymbol{F}}

\newcommand*{\bmW}{\boldsymbol{\mathcal{W}}}

\newcommand*{\indfoc}{\text{foc}}
\newcommand*{\inddiv}{\text{div}}

\newcommand*{\inddet}{\text{det}}
\newcommand*{\indref}{\text{ref}}
\newcommand*{\indcorr}{\text{corr}}
\newcommand*{\inddead}{\text{dead}}
\newcommand*{\indcomp}{\text{comp}}
\newcommand*{\indKL}{\text{KL}}
\newcommand*{\indcal}{\text{cal}}
\newcommand*{\indest}{\text{est}}
\newcommand*{\indpoke}{\text{poke}}
\newcommand*{\norm}[1]{\left\|{#1}\right\|}
\newcommand*{\hooks}[1]{\left [{#1}\right ]}

\newcommand*{\reg}{\mathcal{R}}
\newcommand*{\lovDt}{\lambda/D}
\newcommand*{\ie}{i.e. }

\newcommand*{\review}[1]{{#1}}

\begin{document}

\title{Compensation of high-order quasi-static aberrations on SPHERE with the
  coronagraphic phase diversity (COFFEE)}

\author{B. Paul\inst{1,2,4} \and J.-F. Sauvage\inst{1,4} \and L. M.
  Mugnier\inst{1,4} \and K. Dohlen\inst{2,4} \and C. Petit\inst{1,4} \and T.
  Fusco\inst{1,2,4} \and D. Mouillet\inst{3,4} \and J.-L. Beuzit\inst{3,4}
  \and M. Ferrari\inst{2,4} }

\institute{
Onera - The French Aerospace Lab, F-92322 Ch\^atillon France
\and Aix Marseille Universit\'e, CNRS, LAM (Laboratoire d'Astrophysique de Marseille) UMR 7326, 13388, Marseille, France
\and Institut de Plan\'etologie et d'Astrophysique de Grenoble (IPAG), BP 53 F-38041 Grenoble Cedex 9, France
\and Groupement d'int\'er\^et scientifique PHASE (Partenariat Haute
r\'esolution Angulaire Sol et Espace) between Onera, Observatoire de Paris, 
CNRS, Universit\'e Diderot, Laboratoire d'Astrophysique de Marseille and
Institut de Plan\'etologie et d'Astrophysique de Grenoble}

\date{Received <date>; accepted <date>}

\abstract {The second-generation instrument SPHERE, dedicated to high-contrast
  imaging, will soon be in operation on the European Very Large Telescope.
  Such an instrument relies on an extreme adaptive optics system coupled with
  a coronagraph that suppresses most of the diffracted stellar light.
  However, the coronagraph performance is strongly limited by quasi-static
  aberrations that create long-lived speckles in the scientific image plane,
  which can easily be mistaken for planets. }
{The wave-front analysis performed by SPHERE's adaptive optics system uses a
  dedicated wave-front sensor. The ultimate performance is thus limited by the
  unavoidable differential aberrations between the wave-front sensor and the
  scientific camera, which have to be estimated and compensated for. In this
  paper, we use the COFFEE approach to measure and compensate for SPHERE's
  quasi-static aberrations.}
{COFFEE (for COronagraphic Focal-plane wave-Front Estimation for Exoplanet
  detection), which consists in an extension of phase diversity to
  coronagraphic imaging, estimates the quasi-static aberrations, including the
  differential ones, using only two focal plane images recorded by the
  scientific camera. In this paper, we use coronagraphic images recorded from
  SPHERE's infrared detector IRDIS to estimate the aberrations upstream of the
  coronagraph, which are then compensated for using SPHERE's extreme adaptive
  optics loop SAXO.}
{We first validate the ability of COFFEE to estimate high-order aberrations by
  estimating a calibrated influence function pattern introduced upstream of
  the coronagraph. We then use COFFEE in an original iterative compensation
  process to compensate for the estimated aberrations, leading to a contrast
  improvement by a factor that varies from $1.4$ to $4.7$ between $2\lovDt$
  and $15\lovDt$ on IRDIS. The performance of the compensation process is
  also evaluated through simulations. An excellent match between experimental
  results and these simulations is found.}
{}


\titlerunning{Compensation of quasi-static aberrations on SPHERE with COFFEE}

\authorrunning{B. Paul et al.}

\maketitle

\section{Introduction}

Exoplanet imaging is one of the most challenging areas of today's astronomy.
Such observations, which until now have been only possible for planets with
high masses or wide apparent distances from their host star (\cite{exop_kalas,
  exop_marois, Lagrange-a-09}), can provide information on both the chemical
composition of their atmospheres and their temperatures. The upcoming
ground-based instruments dedicated to exoplanet direct imaging, such as SPHERE
on the VLT (\cite{Beuzit-p-07}) or GPI on Gemini South (\cite{gpi}), will soon
be in operation, providing original data for comparative exoplanetary science
to the community. These instruments rely on extreme adaptive optics (XAO)
systems to ensure a high angular resolution ($0.1'' - 0.1'$) coupled with
coronagraphs to reach the required contrast ($10^6 - 10^7$) on the scientific
detector. The ultimate limitation of these current and future systems lies in
quasi-static aberrations upstream of the coronagraph, which originate in
optical misalignment or surface polishing errors
(\cite{Dohlen-p-11,Hugot-a-12}). These aberrations give birth to long-lived
speckles on the detector, which strongly limit the achievable contrast, since
they can easily be mistaken for a planet. Thus, being able to reach the
ultimate performance of high-contrast imaging systems means estimating and
compensating for these aberrations. The most accurate measurement of these
aberrations can be performed using focal plane wave-front sensors, which are
not limited by non-common path aberrations (NCPA) since they perform the
estimation using data recorded from the scientific camera itself.
 
SPHERE's baseline currently relies on a differential estimation performed with
phase diversity (\cite{Mugnier-l-06a, Sauvage-p-12}), a focal plane wave-front
sensing technique that uses classical imaging (no coronagraph). However, this
wave-front sensor is limited to the estimation of aberrations up to eight
cycles per pupil, which correspond to speckles close to the optical axis (up
to a field angle of $8\lovDt$), whereas the SPHERE XAO system (SAXO) could
compensate for up to $20$ cycles per pupil. Besides, since such a measurement
requires removing the coronagraph, it does not allow proper compensation of
tip, tilt, and defocus aberrations, which code for positioning errors of the
star with respect to the coronagraphic mask.

An optimization of SPHERE's baseline for quasi-static speckle compensation
will thus consist in a focal plane wave-front sensor that retrieves the
aberrations from coronagraphic images, which would allow one to measure
high-order aberrations without removing the coronagraph. Several techniques
dedicated to this goal have been proposed, which all assume small aberrations:
the Self-Coherent Camera (SCC) (\cite{scc}), which relies on a modification of
the imaging system, needs only one image to perform the estimation, whereas
the Electric Field Conjugation (EFC) (\cite{efc}) requires at least three
images to retrieve the aberrations but without any modification of the optical
system.

Our focal plane wave-front sensor, COFFEE (for COronagraphic Focal-plane
wave-Front Estimation for Exoplanet detection), consists in a coronagraphic
extension of phase diversity (\cite{Sauvage-a-12, Paul-a-13a}) that estimates
the aberrations both upstream and downstream of the coronagraph using two
coronagraphic focal plane images. In this paper, we present the application of
the recent high-order myopic extension of this sensor presented in
\cite{Paul-a-13b} to the SPHERE instrument \review{during its final
  integration phase at IPAG (Institut de Planétologie et d'Astrophysique de
  Grenoble). The framework of COFFEE's application to SPHERE is described in
  Section \ref{sect_coffee}}. Section \ref{sect_est} demonstrates the ability
of COFFEE to estimate high-order aberrations with nanometric precision from
experimental focal plane coronagraphic images. Then, Section \ref{sect_comp}
presents SPHERE's contrast optimization (up to $18\lovDt$) on the detector by
compensating for the aberrations (including the high-order ones) using
COFFEE's estimation. Section \ref{sect_ccl} concludes this paper.

\section{Application of COFFEE to SPHERE}
\label{sect_coffee}

COFFEE requires only two images $\bi_c^\indfoc$ and $\bi_c^\inddiv$ recorded
on the detector that differ by a known aberration $\bphi_\inddiv$ to estimate
aberrations both upstream ($\bphi_u$) and downstream ($\bphi_d$) of the
coronagraph. In this paper, we consider the SPHERE instrument calibration,
performed at a high signal-to-noise ratio (SNR) with a monochromatic source
emitted from a single-mode laser fiber. Because it is very small in such a
calibration case, the residual turbulence is therefore neglected in the
sequel. We use the following coronagraphic imaging model:
\begin{equation}\label{eq_im_model}
\begin{aligned}
\bi_c^\indfoc&=\alpha_\indfoc \bh_\inddet \star \bh_c(\bphi_u,\bphi_d)+\bn_\indfoc+\beta_\indfoc\\
\bi_c^\inddiv&=\alpha_\inddiv \bh_\inddet \star \bh_c(\bphi_u+\bphi_\inddiv,\bphi_d)+\bn_\inddiv+\beta_\inddiv
\end{aligned}
\end{equation}
where $\alpha_p$ is the incoming flux ($p$ stands for ``foc'' or ``div''),
$\bh_c$ the coronagraphic ``point spread function'' (PSF) of the instrument
(which depends on $\bphi_u$ and $\bphi_d$, and whose expression is explained
in \cite{Paul-a-13b}), $\bh_\inddet$ the known detector PSF, $\bn_\indfoc$ and
$\bn_\inddiv$ are the measurement noises that comprise both detector and
photon noises, $\beta_p$ is a unknown uniform background (offset), and $\star$
denotes the discrete convolution operation. COFFEE estimates the aberrations
$\bphi_u$ and $\bphi_d$, as well as the fluxes
$\balpha=[\alpha_\indfoc,\alpha_\inddiv]$ and the backgrounds
$\bbeta=[\beta_\indfoc,\beta_\inddiv]$ that minimize the maximum \textit{a
  posteriori} (MAP) criterion $J(\balpha, \bbeta, \bphi_u, \bphi_d )$ whose
expression is given hereinafter.

\begin{figure}
\centering
\includegraphics[width = 1\linewidth]{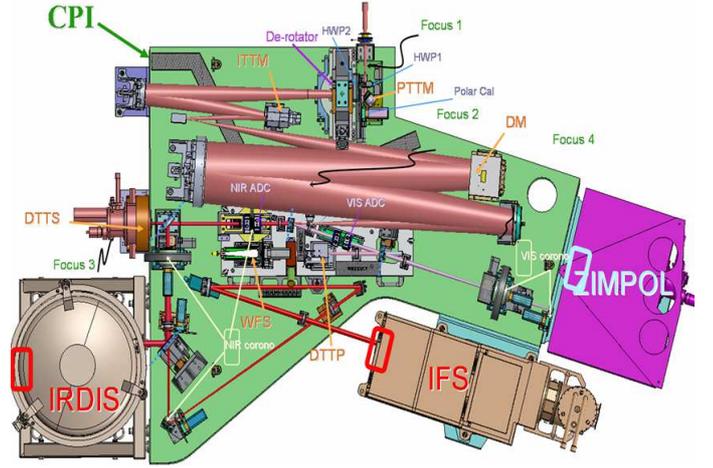}
\caption{\review{Schematic representation of SPHERE. To apply COFFEE on this
  instrument, coronagraphic images were recorded from the IRDIS detector.}}
\label{sphere_ov}
\end{figure}

\begin{figure}
\centering
\begin{tabular}{cc}
\includegraphics[width = 0.45\linewidth]{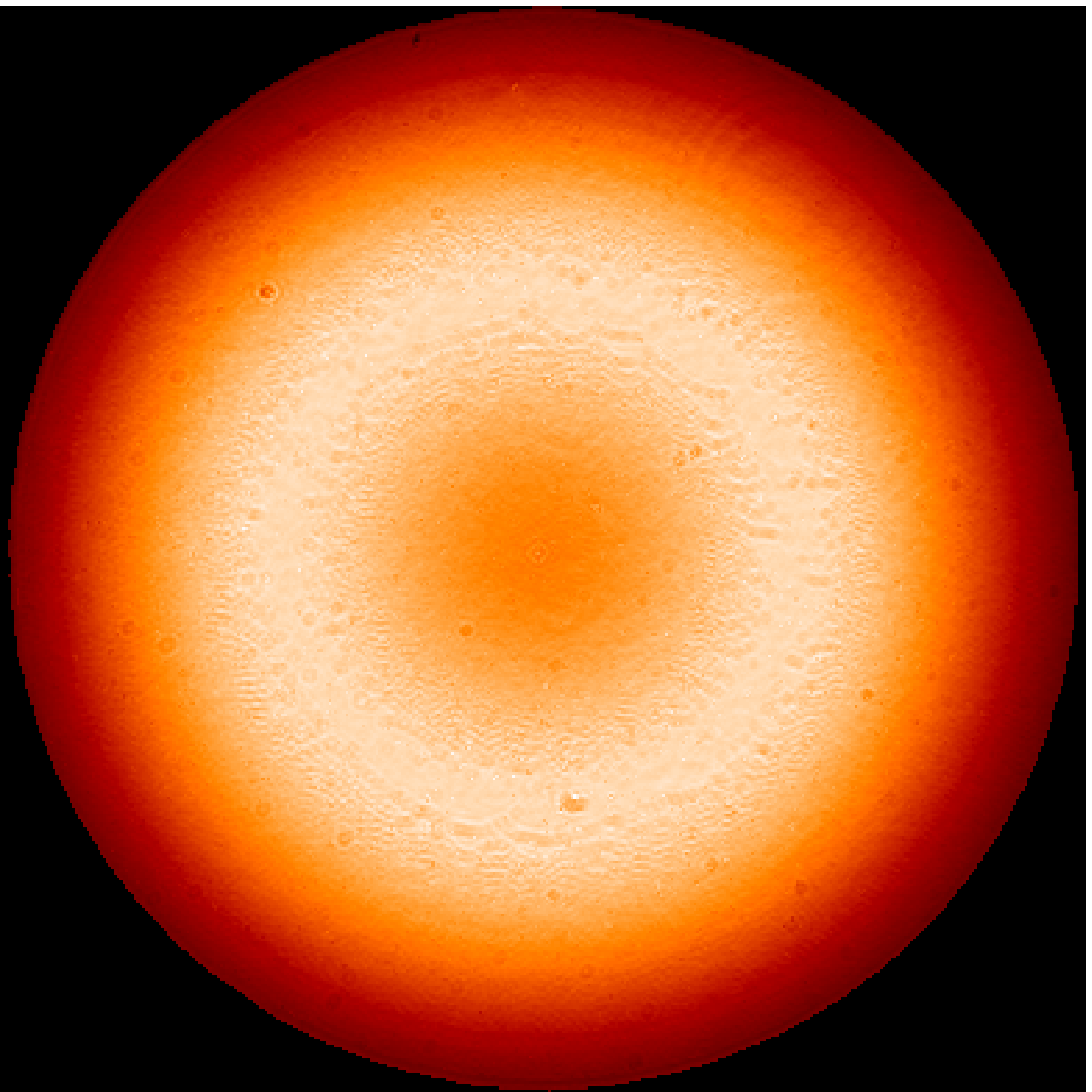}&
\includegraphics[width = 0.45\linewidth]{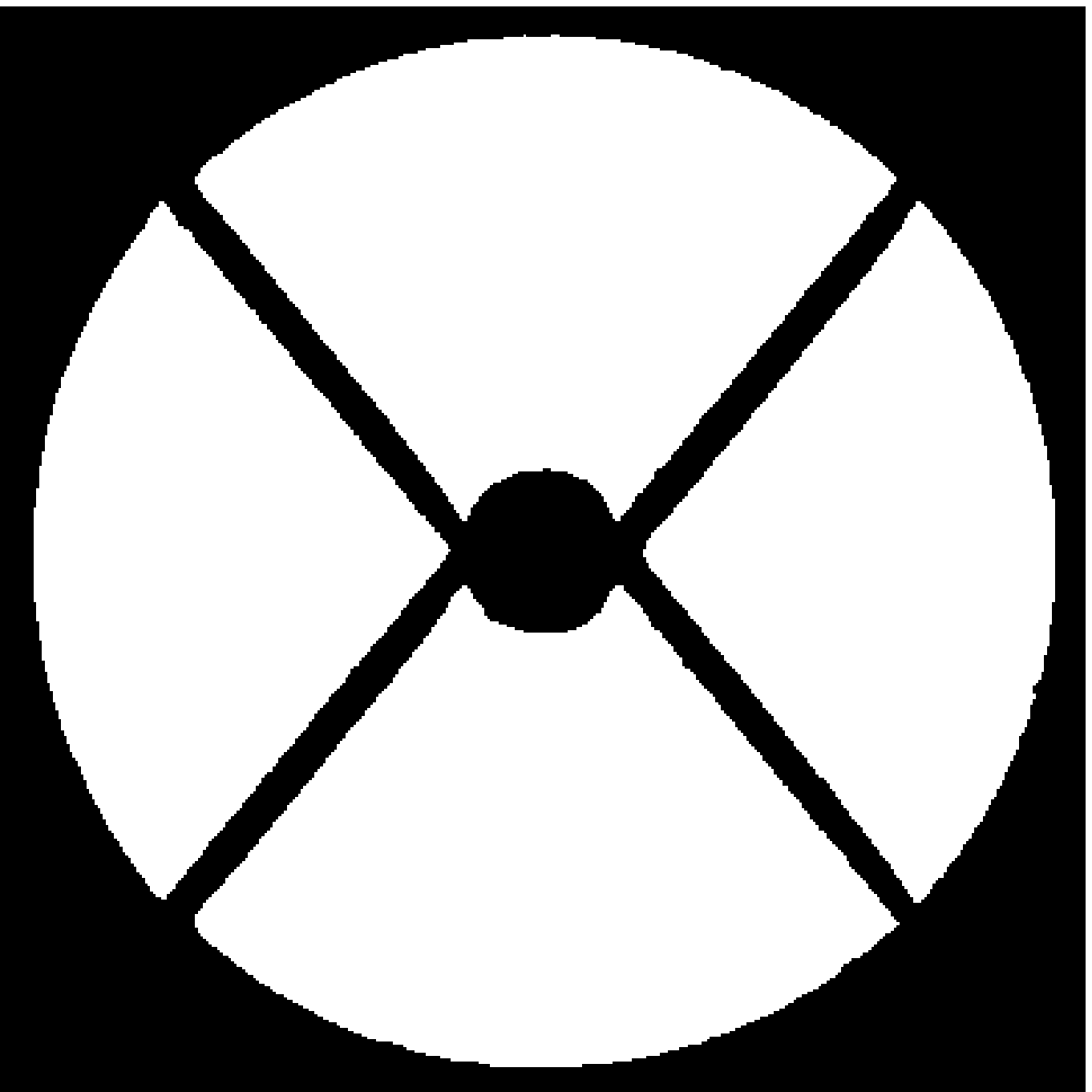}\\
\end{tabular}
\caption{Experimental images of SPHERE's entrance pupil apodizer (left) and of
  the Lyot Stop pupil transmission (right).}
\label{apod_lyot}
\end{figure}

\review{The SPHERE system is represented in figure \ref{sphere_ov}. To use
  COFFEE on this instrument, coronagraphic images are recorded by the
  infrared camera IRDIS (infrared dual imaging spectrograph) using an
  apodized Lyot coronagraph (ALC), whose apodizer is presented on Figure
  \ref{apod_lyot} (left), and an obscurated ($15\%$) Lyot Stop pupil whose
  transmission is presented on figure \ref{apod_lyot} (right) as well. All the
  parameters used for coronagraphic image acquisitions are gathered in Table
  \ref{table_param_sphere}. The SPHERE XAO loop SAXO will be described in
  section \ref{sect_comp_th}.}

\begin{tiny}
\begin{table}
\centering
\begin{tabular}{m{3cm} m{5cm}}
  \hline
  \review{Light source} & \review{Internal calibration source, wavelength} $\lambda = 1589$ \review{nm}\\
  \review{SPHERE entrance pupil} & \review{Unobscured circular pupil, diameter} $D_u = 40$ \review{cm}\\
  \review{Coronagraph} & \review{Apodized Lyot Coronagraph (ALC), focal plane mask angular diameter} $d=4.52\lovDt$\\  
  \review{Lyot stop pupil} & \review{Obscured circular pupil (see figure
    \ref{apod_lyot}), diameter} $D_d = 0.96D_u$\\
  \review{IRDIS image size} & $200\times 200$ \review{pixels}\\
  \review{Sampling} & $2.75$ \review{pixels per} $\lovDt$\\
  \review{Detector noise} & $\sigma_\inddet=1\ \text{e}^-$\\
  \review{Exposure time} & $0.6$ \review{seconds per acquisition. Each image is
    averaged over $100$ acquisitions}\\
  \hline
\end{tabular}
\caption{\review{Parameters used to apply COFFEE on SPHERE}}
\label{table_param_sphere}
\end{table}
\end{tiny}

It is worth mentioning that several ``dead'' pixels can be found on the IRDIS
CCD detector (around $1\%$ per image). Such pixels, whose value is notably
high, could strongly limit COFFEE's accuracy by introducing a bias in
criterion $J$'s value. Thus, to improve COFFEE's accuracy, these
pixels are detected in each recorded image prior to the phase estimation and
taken into account by modifying criterion $J$'s expression given in
\cite{Paul-a-13b} as follows:
\begin{equation}\label{eq_pb_inverse}
\begin{aligned}
J&=\frac{1}{2} \norm{\bmW^\indfoc\hooks{\bi_c^{\text{foc}} 
- (\alpha_\indfoc \bh_\inddet\star \bh_c(\bphi_u,\bphi_d)+\beta_\indfoc)}}^2 \\
&+\frac{1}{2}\norm{\bmW^\inddiv\hooks{\bi_c^{\text{div}} 
- (\alpha_\inddiv \bh_\inddet\star \bh_c(\bphi_u+\bphi_\inddiv,\bphi_d)+\beta_\inddiv)}}^2\\
&+\reg(\bphi_u) + \reg(\bphi_d)\text{,}
\end{aligned}
\end{equation}
where $\norm{x}^2$ denotes the sum of squared pixel values of map $x$. Here,
$\bmW^\indfoc$ and $\bmW^\inddiv$ are ``weight'' maps defined as $\bmW=0$ if
the pixel is detected as dead and $\bmW=1/\bsigma_n^2$ otherwise, where
$\bsigma_n^2$ is the noise variance in the image. Thus, by setting an infinite
variance on this pixel (\ie $\bmW^p=0$), dead pixels no longer have an impact
on the criterion value, and thus on COFFEE's estimation.

As explained in \cite{Paul-a-13b}, regularization metrics $\reg(\bphi_k)$
(where $k$ stands for $u$ (upstream) or $d$ (downstream)) used in criterion
$J$ (Eq.~\eqref{eq_pb_inverse}) are based on the available \textit{a priori}
knowledge on the SPHERE quasi-static aberration's power spectral density
(PSD), which follows a $1/\nu^{2}$ scaling law, where $\nu$ is the spatial
frequency (\cite{Dohlen-p-11,Hugot-a-12}). These metrics are given by
\begin{equation}
\reg(\bphi_k)=\frac{1}{2\sigma_{\nabla\bphi_k}^2}\norm{\nabla\bphi_k}^2\text{,}
\end{equation}
where $\nabla$ represents the gradient operator, and $\sigma_{\nabla\bphi_k}^2$
the variance of $\nabla\bphi_k$, which can be computed from the
aberration's PSD and the RMS value of the wave-front error (WFE).

The diversity phase $\bphi_\inddiv$ used to record the diversity image
$\bi_c^\inddiv$, introduced using SAXO, is composed of defocus alone
($\bphi_\inddiv = a_\inddiv Z_4$). The amplitude $a_\inddiv$ has been chosen
following \cite{Paul-a-13b} where it has been demonstrated that if a pure
defocus is used as diversity phase, its amplitude should be $2.5$ times greater
than the WFE of the aberration upstream of the coronagraph
for an optimal estimation. Knowing from SPHERE's baseline phase diversity
measurement that the WFE upstream of the coronagraph is approximately $40$ nm
RMS ($0.16$ rad RMS), we use $a_\inddiv=101$ nm RMS ($0.4$ rad RMS).

\section{High-order aberration estimation}
\label{sect_est}

\begin{figure}
\centering
\begin{tabular}{cc}
\includegraphics[width = 0.45\linewidth]{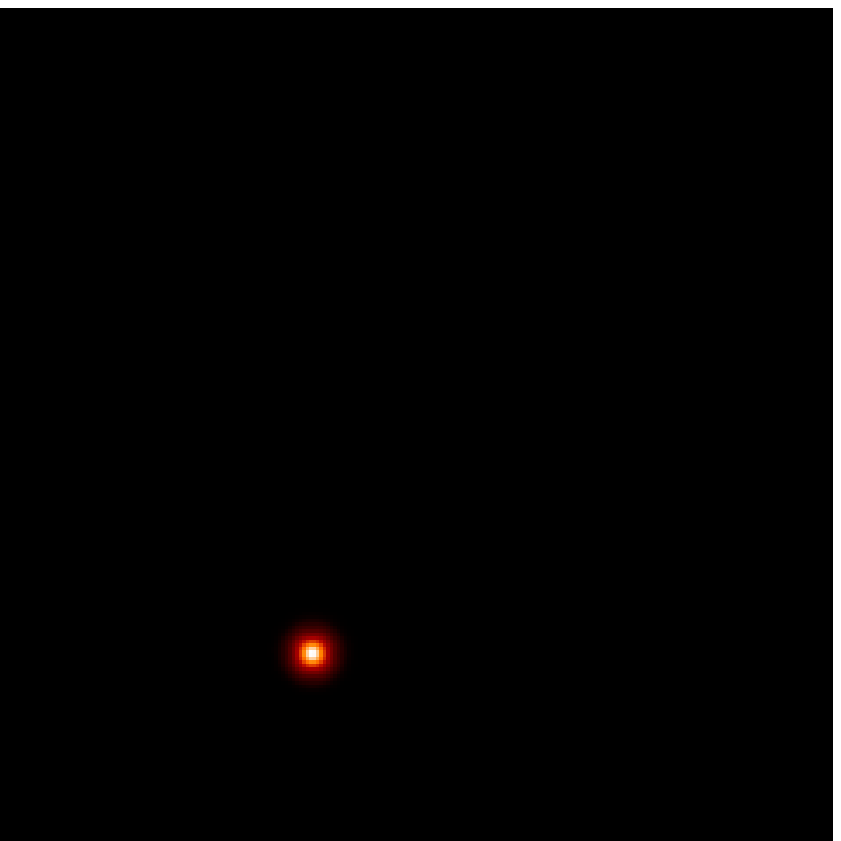}&
\includegraphics[width = 0.45\linewidth]{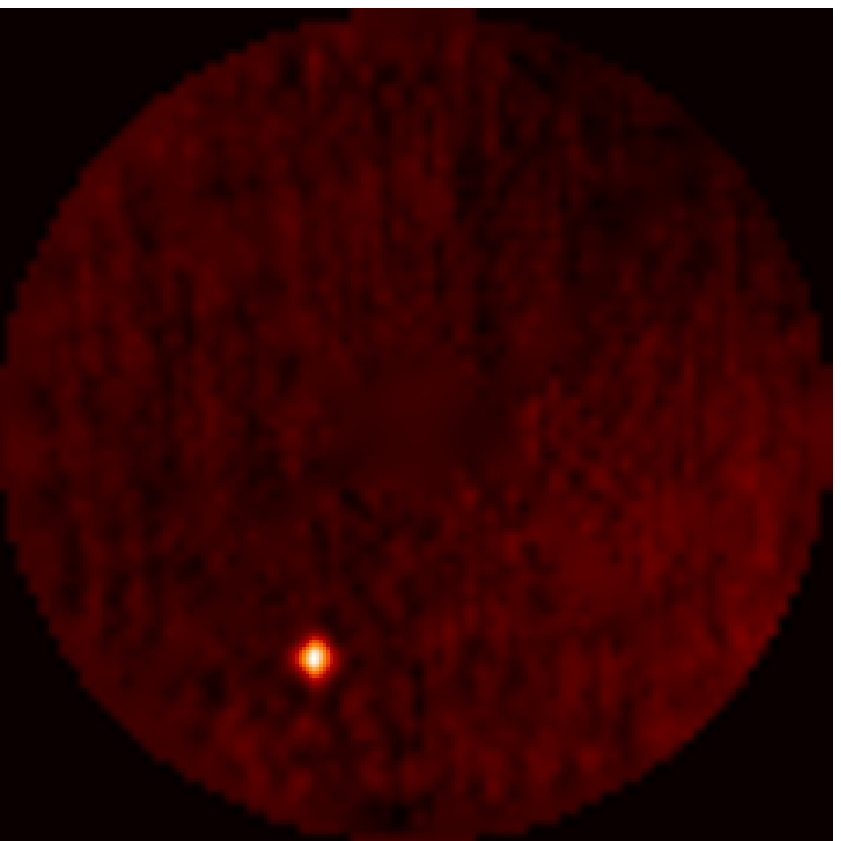}\\
\includegraphics[width = 0.45\linewidth]{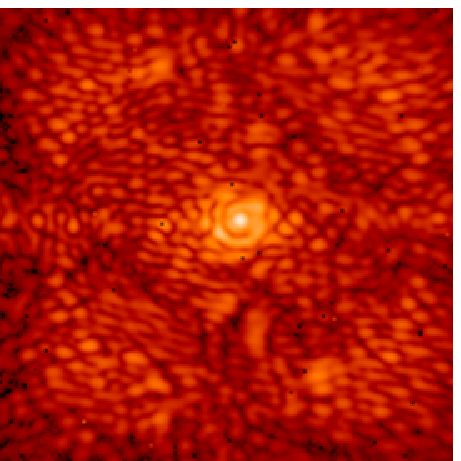}&
\includegraphics[width = 0.45\linewidth]{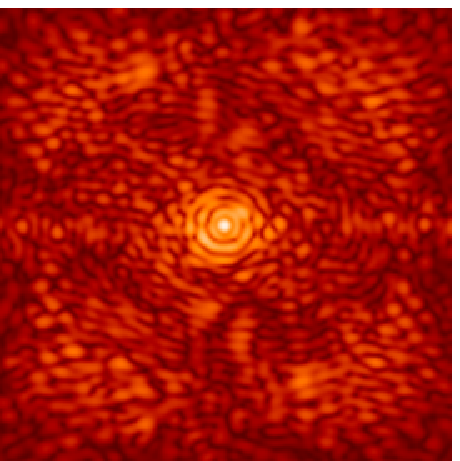}\\
\end{tabular}
\caption{High-order aberration (poke) estimation on SPHERE with COFFEE. Top:
  introduced poke (left, $\epsilon=144$ nm PV) and COFFEE-estimated poke
  (right, $\epsilon_\indest=147$ nm PV). Bottom: experimental image recorded from
  IRDIS (left) and image computed using the reconstructed aberration (right)
  (log. scale, same range for both images).}
\label{est_poke}
\end{figure}

We first validate COFFEE on SPHERE by estimating a high-order calibrated
aberration introduced upstream of the coronagraph. First, focused and
diversity images are recorded from IRDIS, allowing COFFEE to estimate SPHERE's
aberrations $\hbphi_u^{\indref}$. Then, by pushing on a single actuator
(amplitude $\epsilon_\indcal=144$ nm PV) of SAXO's high-order deformable
mirror (HODM), we introduce the corresponding influence function pattern
(called hereafter a poke) upstream of the coronagraph. \review{This poke
  pattern is represented in the top left of figure \ref{est_poke}}. Focused
and diversity coronagraphic images are recorded from IRDIS and then processed
by COFFEE, which estimates an aberration $\hbphi_u^{\indpoke}$ upstream of the
coronagraph. Then, the difference
$\bphi_u^{\indest}=\bphi_u^{\indpoke}-\bphi_u^{\indref}$ gives the poke
estimated by COFFEE.
As one can see at the top of Figure \ref{est_poke}, COFFEE's estimated poke is
very close to the introduced one: the difference between the two is $3$ nm PV.
This results, in turn, in a very good match between the experimental images
and the ones computed for the estimated aberrations (Figure \ref{est_poke},
bottom). From COFFEE's estimation (Figure \ref{est_poke}, top right), one can
notice, apart from the poke, a residual high-frequency aberration whose
amplitude is $6.82$ nm RMS, representing 5\% of the poke value. Its origin
lies in a combination of several terms, including internal turbulence, DM
crosstalk, and computational effects.

\section{Contrast optimization on IRDIS: quasi-static aberration compensation
  with COFFEE}
\label{sect_comp}

In this section, we propose a refined process for the compensation of SPHERE's
quasi static aberrations using the phase map $\hbphi_u^{\indref}$ estimated
by COFFEE. Section \ref{sect_comp_th} presents the application of the
pseudo-closed loop process (PCL) proposed in \cite{Paul-a-13a} to SPHERE,
which is then used in Section \ref{sect_comp_exp} to optimize the contrast on
IRDIS. The obtained performance of the PCL is then discussed in Section
\ref{sect_comp_sim} and cross-validated using simulations. Last, in Section
\ref{sect_comp_psd}, we analyze the aberrations upstream of the coronagraph
estimated by COFFEE before and after compensation.

\subsection{Projection of the estimated aberrations on SAXO's reference
  slopes}
\label{sect_comp_th}

After their estimation, the aberrations upstream of the coronagraph $\hbphi_u$
are transformed into a set of slopes that will then be used to modify SAXO's
references slopes $\bs_\indref$, following the PCL process described in
\cite{Paul-a-13a}. 

We let $\bF$ be the HODM calibrated influence matrix; any HODM introduced
aberration can be described as a set of $1377$ actuator voltages
$\bv_\indcorr$. We are thus looking for the set $\bv_\indcorr$ that solves
the least-squares problem:
\begin{equation}
\bv_\indcorr=\underset{\bv}{\arg\min}\norm{\bF\bv - \hbphi_u}^2.
\end{equation}
The solution of this problem can be written as
\begin{equation}
  \bv_\indcorr=\bT\hbphi_u\text{,}
\end{equation}
with $\bT$ the generalized inverse of matrix $\bF$. 

Using the calibrated Shack-Hartmann wave-front interaction matrix $\bD$, we
can compute the corresponding set of slopes $\bs_\indcorr=\bD\bv_\indcorr$.
Thus, the simplest way to compensate for the aberration $\bphi_u$ upstream of
the coronagraph would consist in introducing the estimated aberration
$\hbphi_u$ with SAXO by modifying the references slopes $\bs_\indref$ of the
wave-front sensor,
\begin{equation}\label{eq_pcl_pcp}
\bs_\indcomp = \bs_\indref -g\bs_\indcorr\text{,}
\end{equation}
where $g$ is the PCL gain and $\bs_\indcomp$ are SAXO's updated references
slopes.

We then denote by $\bC$ the SAXO matrix that controls the HODM by computing,
from the slopes $\bs_\indcomp$, the corresponding set of voltages
$\bv_\indcomp=\bC\bs_\indcomp$. The conventional way to obtain this matrix is
to compute it as the generalized inverse of $\bD$. Various modal bases can be
used to control the HODM. The simplest one corresponds to the eigen modes of
the system, computed from $\bD$'s inversion. SAXO's robustness can be improved
by truncating the control basis, which is conventionally performed by
filtering out the modes that correspond to low eigenvalues from the basis.
However, \cite{Petit-p-08a} demonstrate that because of the large number of
HODM actuators, some of these low eigenvalues correspond, in the case of SAXO,
to high-energy modes whose truncation from the control basis would lead to a
significant performance decrease.

Thus, \cite{Petit-p-08a} have determined that an optimized control of the HODM
can be performed with a Karhunen--Loeve (KL) control basis computed on the
space defined by the HODM influence functions. The $1377$ KL modes of this
basis, naturally ordered according to the propagated turbulent energy, allow a
proper control basis truncation. In SAXO's case, a robust and efficient
control of the HODM can be achieved using a $999$ modes control basis. It is
worth mentioning that such a basis takes the HODM actuators into account that
cannot be controlled by the loop, either because they will be located under
the telescope obscuration or because the are optically or electronically dead.
Thus, to accurately introduce the aberration $\hbphi_u$ on SPHERE using SAXO,
it is necessary to modify Equation \eqref{eq_pcl_pcp}, where the slopes
$\bs_\indcorr$ are computed considering that the $1377$ HODM actuators are
controlled. To accurately introduce $\hbphi_u$ using SAXO, it is indeed
necessary to compute the corresponding slopes $\bs_\indcorr^\indKL$ that
modify only the $999$ KL modes controlled by the loop. The available matrices
used by SAXO allow us to compute the matrix $\bM$ that describes the slopes
$\bs_\indcorr$ as a set of $999$ KL modes $\bm_\indcorr=\bM\bs_\indcorr$ and
$\bN$ its generalized inverse. Thus, $\bN\bM$ is the projection that allows
computing the slopes $\bs_\indcorr^\indKL=\bN\bM\bs_\indcorr$ which are then
used to modify Equation \eqref{eq_pcl_pcp}:
\begin{equation}\label{eq_pcl_saxo}
\begin{aligned}
\bs_\indcomp &= \bs_\indref -g\bs_\indcorr^\indKL\\
&=\bs_\indref -g\bN\bM\bD\bT\hbphi_u.
\end{aligned}
\end{equation}

The PCL compensation process on SAXO described above can thus be described as
follows. At iteration $j$, SAXO is closed on a set of reference slopes
$\bs_\indref^j$:
\begin{enumerate}
\setlength\itemsep{-0.1in}
\item acquisition of the focused $\bi_c^\indfoc$ and diverse
    $\bi_c^\inddiv$ images with IRDIS;\\
\item estimation of the aberration $\hbphi_u^j$ upstream of
  the coronagraph using these images with COFFEE;\\
\item computation of the corresponding reference slopes correction:
  $\delta\bs_\indcorr^\indKL=\bN\bM\bD\bT\hbphi_u^j$;\\
\item modification of SAXO's reference slopes whose computation is
  given by equation \eqref{eq_pcl_saxo}.
\end{enumerate}

Thus, the aberrations $\bphi_u^{j+1}$ upstream of the coronagraph at the
iteration $j+1$ of the PCL, computed from COFFEE's estimated aberration
$\hbphi_u^j$, can be written as
\begin{equation}\label{eq_pclsaxophi}
\bphi_u^{j+1} = \bphi_u^j - g\bF\bC\bN\bM\bD\bT\hbphi_u^j\text{.}
\end{equation}

\subsection{Contrast optimization on SPHERE}
\label{sect_comp_exp}

\begin{figure}
\centering
\begin{tabular}{cc}
\includegraphics[width = 0.4\linewidth]{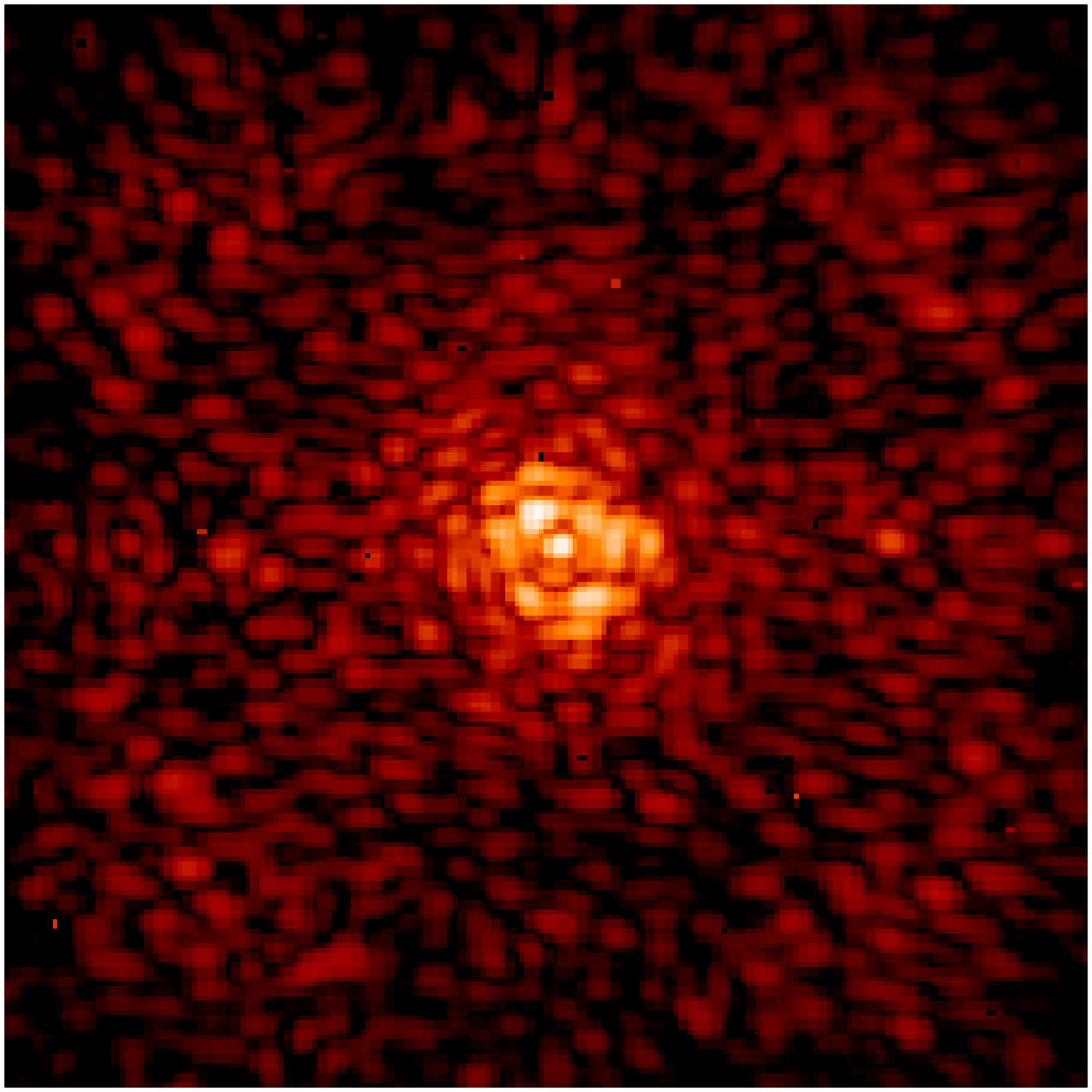}&
\includegraphics[width = 0.4\linewidth]{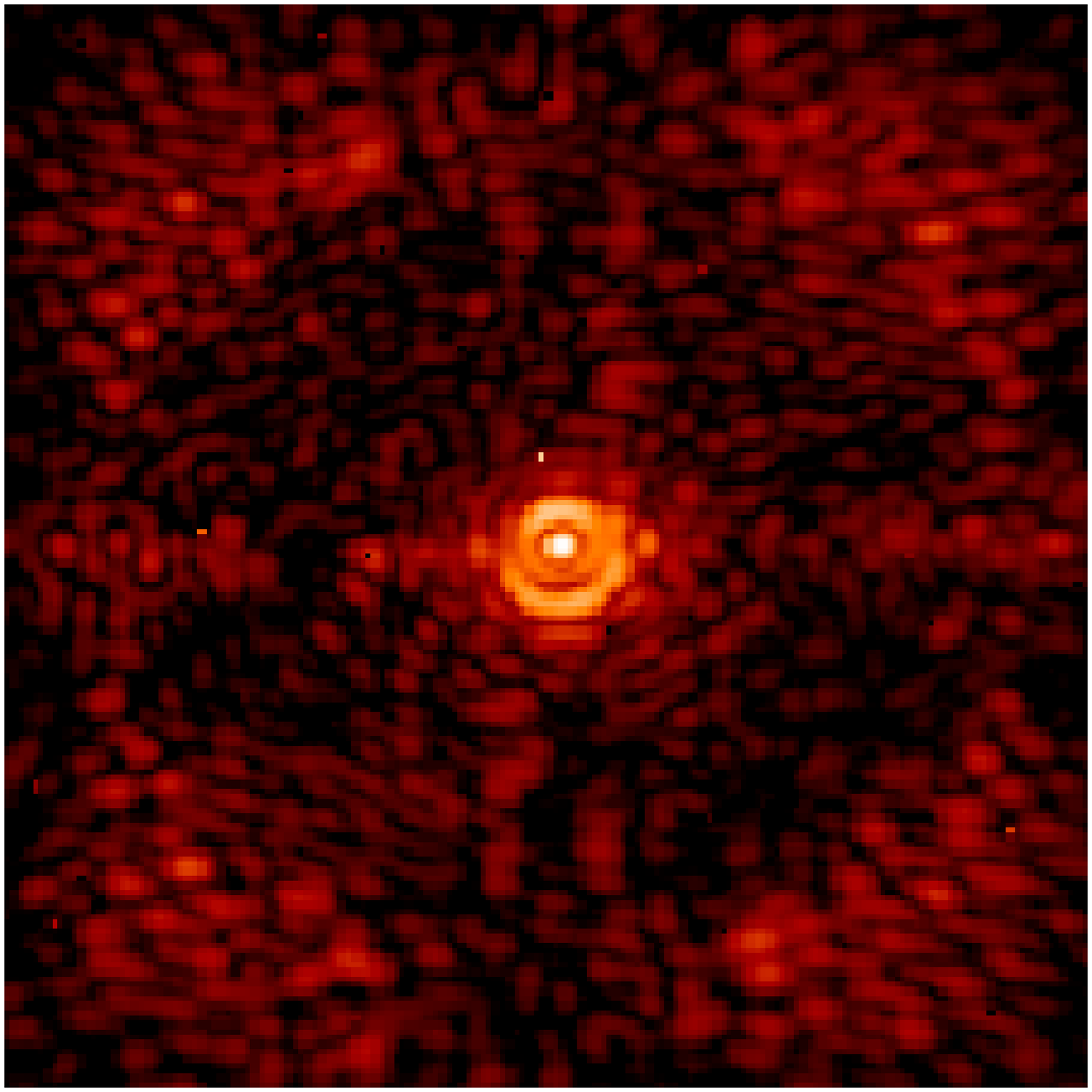}\\
\multicolumn{2}{c}{\includegraphics[width = 0.95\linewidth]{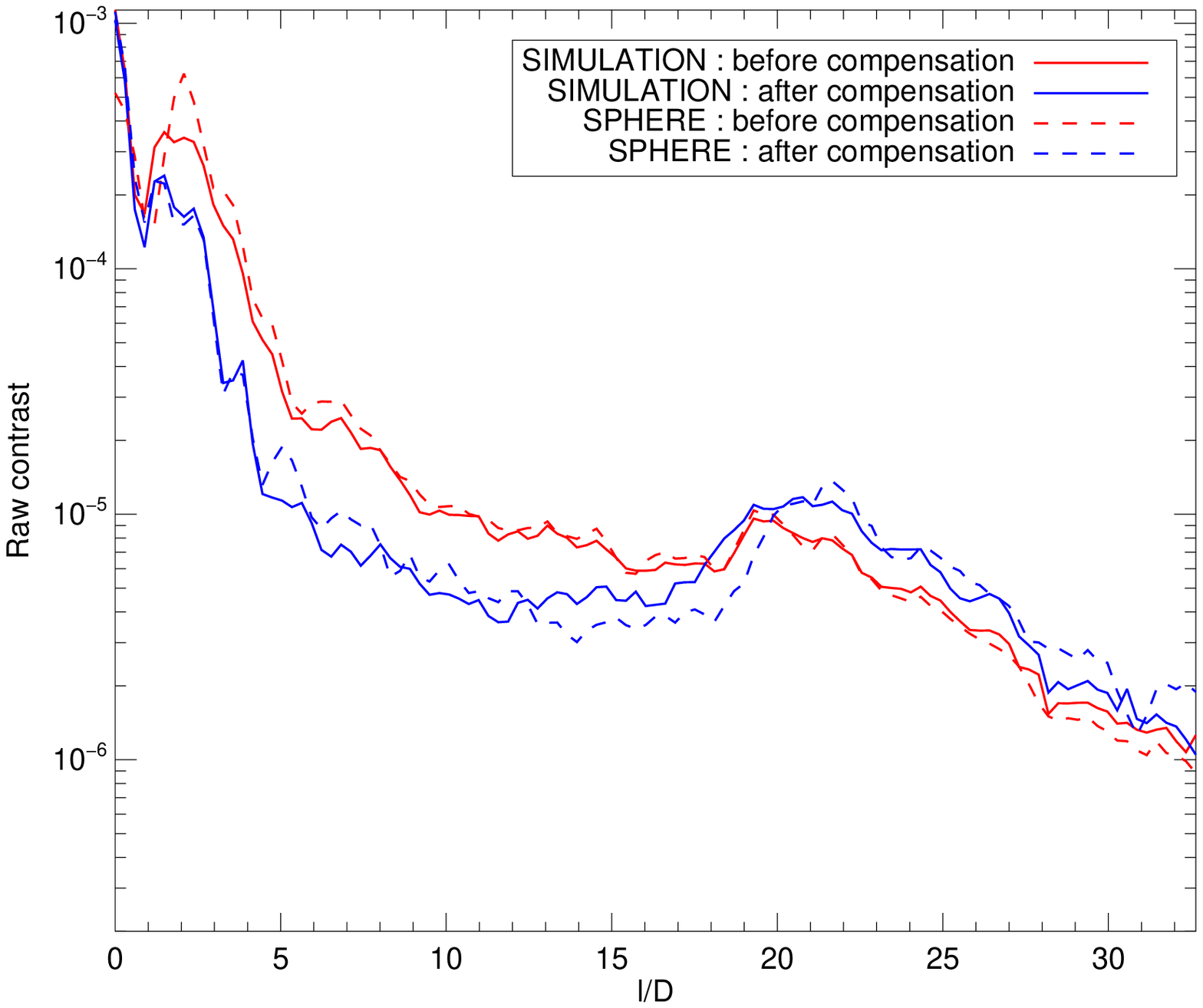}}\\
\end{tabular}
\caption{Aberration compensation with COFFEE: PCL on SPHERE ($g=0.5$). Top:
  coronagraphic images recorded from IRDIS before compensation (left) and
  after $5$ PCL iterations (right. log. scale, same dynamic for both images).
  Bottom: averaged raw contrast profiles computed from the experimental images
  before (dashed red line) and after (dashed blue line) compensation. For
  comparison, raw contrast profiles computed from a complete simulation of
  SAXO before (solid red line) and after compensation (solid blue line) are
  plotted}
\label{exp_cp}
\end{figure}

Figure \ref{exp_cp} shows the result of the PCL process on SPHERE. The average
computation time (Step 2) on a standard PC is $\Delta t=9.7$ minutes for one
iteration, allowing us to compensate for SPHERE quasi-static aberrations.
Indeed, the SPHERE quasi-static WFE has been found to increase at a rate of
$\Delta\text{WFE}=0.07$ nm RMS in $H-$band per minute (\cite{Martinez-a-13}),
which results in an estimation error of $\Delta t\times\Delta\text{WFE}=0.7$
nm RMS, which can be neglected.

In the image recorded by IRDIS after five PCL iterations (Figure \ref{exp_cp},
top), the aberration compensation has obviously removed speckles in most of
the detector plane area controlled by SAXO. The corresponding contrast gain is
illustrated by the profiles plotted at the bottom of Figure \ref{exp_cp}
(dashed lines). In particular, the contrast is improved by a factor that
varies from $1.4$ to $4.7$ between $2\lovDt$ and $15\lovDt$. We note that such
a gain is superior to the one expected from SPHERE's baseline (classical phase
diversity), which in addition, does not improve the contrast beyond $8\lovDt$
(\cite{Sauvage-p-12}). \review{The origin of the contrast gain limitation
  observed here on the IRDIS camera is thoroughly studied in the next
  section.}

Comparing the averaged contrast profiles computed after compensation, one can
notice that beyond $18\lovDt$, the intensity increases on the detector. We
have determined and checked by simulations that this behavior is due to the
DM's central actuators, which are not controlled in the same way as the other
actuators, since they will be masked by the telescope central obscuration
during on-sky observations. \review{However, such an obscuration is not
  present in the entrance pupil during the calibration phase, hence the energy
  increase that can be observed in Figure \ref{exp_cp}.} It is worth
mentioning that this increase in high-order speckles does not appear when an
obscuration is present in the entrance pupil, as will be the case during
scientific observations.

\subsection{Compensation process: performance assessment}
\label{sect_comp_sim}

This section aims at assessing the performance of the PCL process described in
Section \ref{sect_comp_th}. This assessment is performed using a simulation
that closely mimics the SPHERE instrument. The aberrations $\bphi_u$ used to
simulate coronagraphic images before compensation are the ones estimated by
COFFEE from IRDIS images; besides, we include an inhomogenehous entrance pupil
transmission (amplitude aberration), extracted from a entrance pupil plane
experimental image recorded from IRDIS. The simulated PCL process closely
follows the one used on SPHERE and described in Section \ref{sect_comp_th}:
\begin{enumerate}
\setlength\itemsep{-0.1in}
\item at iteration $j$, computation of the coronagraphic focused
  $\bi_c^\indfoc$ and diverse $\bi_c^\inddiv$ images using the image
  formation model described in Equation \eqref{eq_im_model};\\
\item estimation of the aberrations $\hbphi_u^j$ upstream of
  the coronagraph using these images;\\
\item computation of the aberrations upstream of the coronagraph
  $\hbphi_u^{j+1}$ at iteration $j+1$ using Equation \eqref{eq_pclsaxophi}.
\end{enumerate}

One can note here that the computation of $\hbphi_u^{j+1}$, performed with all
SAXO's matrices, allows us to accurately include SPHERE's AO loop in this PCL
simulation. In particular, we take here the presence of the actuators
uncontrolled by SAXO into account (such as the dead actuators mentioned in
Section \ref{sect_comp_th}), as well as the fact that SAXO controls $999$ KL
modes instead of the $1377$ HODM actuators. As presented in Figure
\ref{exp_cp}, this results, in turn, in an excellent match between the
contrast profiles computed from simulation before and after compensation and
the one computed from experimental images recorded from IRDIS.

To evaluate the impact of SAXO's limitations on the PCL process, we
simulated \review{two other scenarii of PCL process} with a simplified
compensation stage (step $3$ of the simulated PCL). First, we considered a
compensation process where the aberration $\bphi_u^{j+1}$ upstream of the
coronagraph at the iteration $j+1$ are computed as
\begin{equation}\label{eq_pclth}
\bphi_u^{j+1} = \bphi_u^j - g\bF_0\bT_0\hbphi_u^j\text{,}
\end{equation}
where $\bF_0$ is the influence matrix that corresponds to a perfect $1377$
actuators DM, \ie a DM where all actuators can be controlled, and $\bT_0$ its
generalized inverse. The compensation described in Eq.~\eqref{eq_pclth} can
thus be considered as an ideal AO loop that would control the $1377$ HODM
actuators perfectly, \ie without any dead actuators or control basis
truncation.

\review{Besides}, in order to distinguish between the contrast limitation due
to dead actuators and the one that comes from \review{the KL modes that are
  filtered out to improve SAXO's robustness}, a \review{third scenario of} PCL
was simulated. In this compensation process, where the only limitation lies in
dead actuators, the aberration $\bphi_u^{j+1}$ upstream of the coronagraph at
the iteration $j+1$ are computed as
\begin{equation}\label{eq_pcldead}
\bphi_u^{j+1} = \bphi_u^j - g\bF_\inddead\bT_\inddead\hbphi_u^j\text{.}
\end{equation}
Here, $\bF_\inddead$ denotes the influence matrix of a $1377$-actuator DM
where the eight influence-function patterns that correspond to SAXO's dead
actuators have been set to $0$. Thus, this compensation process allows
simulating an AO loop without any control basis truncation, where all
actuators but the dead ones are controlled.

\begin{figure}
\centering
\includegraphics[width = 0.95\linewidth]{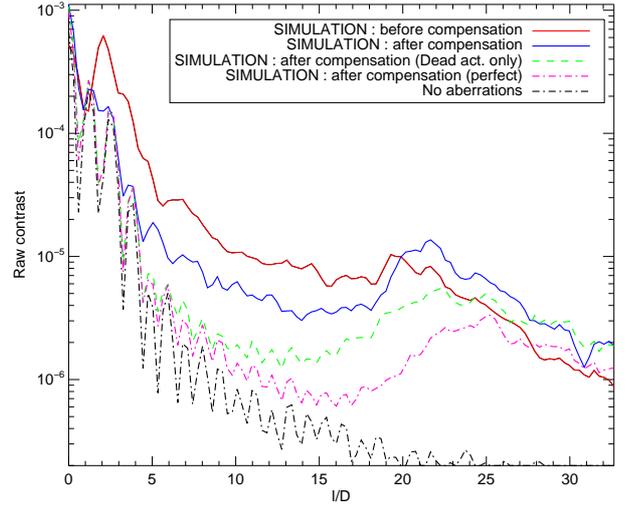}
\caption{PCL simulation in a SPHERE-like case ($g=0.5$): raw contrast profiles
  computed before (solid red line) and after compensation with a complete
  simulation of SAXO ($C_{\text{SAXO}}$, solid blue line), without KL modes
  filtered out ($C_\inddead$, dashed green line), and considering an ideal AO
  loop (i.e., without dead actuators or filtered out modes, $C_{\text{AO}}$,
  dotted dashed magenta line). For comparison, the contrast profiles computed
  from a perfect coronagraphic image (no aberrations) is plotted (dotted
  dashed black line).}
\label{sim_cp}
\end{figure}

\begin{tiny}
\begin{table}
\centering
\begin{tabular}{m{1.5cm} m{1.9cm} m{1.9cm} m{1.9cm}}
\hline
$d$ & $C_{\text{SAXO}}$ & $C_\inddead$ & $C_{\text{AO}}$ \\
\hline
$5 \lovDt$ & $1.9\ 10^{-5}$ & $7.2\ 10^{-6}$ & $5.9\ 10^{-6}$ \\
$10 \lovDt$ & $6.2\ 10^{-6}$ & $1.9\ 10^{-6}$ &  $1.4\ 10^{-6}$ \\
$15 \lovDt$ & $3.8\ 10^{-6}$ & $1.6\ 10^{-6}$ &  $7.6\ 10^{-7}$ \\
\hline
\end{tabular}
\caption{PCLs simulation: Contrast value computed after a compensation
  performed with a complete simulation of SAXO ($C_{\text{SAXO}}$)\review{,
    with a compensation limited only by dead actuators ($C_\inddead$)}, and
  with an ideal AO loop ($C_{\text{AO}}$) at different distances from the
  axis in the coronagraphic images.}
\label{tab_cval}
\end{table}
\end{tiny}

In Figure \ref{sim_cp}, the comparison of the contrasts obtained with a
complete simulation of SAXO $C_{\text{SAXO}}$ and the one obtained with an
ideal AO loop $C_{\text{AO}}$ \review{whose values are given} in Table
\ref{tab_cval} \review{at different positions on the focal plane} clearly
demonstrates that the main limitation of the PCL process performed on SPHERE
lies in SAXO's control law. Indeed, because of \review{the truncation of its
  control basis}, SAXO cannot compensate for the aberrations that give birth
to the remaining speckles. \review{Besides, from the contrast profile
  computed} with an AO loop limited only by SAXO's dead actuators (Figure
\ref{sim_cp}), \review{one can see} that these dead actuators limit the
achievable contrast mainly far from the optical axis (beyond $15\lovDt$).
\review{Such behavior demonstrates that the} difference between this contrast
profile and the one computed with a complete simulation of SAXO, which is
especially important below $15\lovDt$, originates in the \review{other
  filtered-out KL modes of the control basis.}

\review{Thus, the simulations presented in this section demonstrate that the
  PCL process used in SPHERE in Section \ref{sect_comp_exp} is limited by
  SAXO's control basis truncation. Indeed, dead actuators limit the achievable
  contrast far from the optical axis (beyond $15\lovDt$), whereas the others
  KL modes strongly limit the contrast between $0\lovDt$ and $15\lovDt$, when
  filtered out to improve SAXO's robustness.}

Finally, we note that although they cannot be accurately compensated with
SAXO, all aberrations upstream of the coronagraph are still accurately
estimated by COFFEE. Indeed, one can see that with an ideal AO loop
(Eq.~\eqref{eq_pclth}), the contrast computed in the focal plane after
compensation using COFFEE (Figure \ref{sim_cp}) is very close to the one that
is computed from a coronagraphic PSF simulated without any phase or amplitude
aberration (Figure \ref{sim_cp}). This behavior demonstrates the aberration
estimation accuracy, which given a perfect compensation stage, would allow one
to almost reach the theoretical performance offered by the considered
coronagraphic device. \review{These results demonstrate in particular that
  amplitude aberrations, represented in these simulations by the entrance
  pupil inhomogeneous transmission, have a very small impact on the achievable
  contrast in the focal plane, at least in the SPHERE framework. Indeed, the
  contrast computed with the ALC coronagraph without any aberrations can
  almost be reached even though amplitude aberrations are neglected.}

\subsection{Analysis of SPHERE residual aberrations}
\label{sect_comp_psd}

\begin{figure}
\centering
\begin{tabular}{c}
\includegraphics[width = 0.8\linewidth]{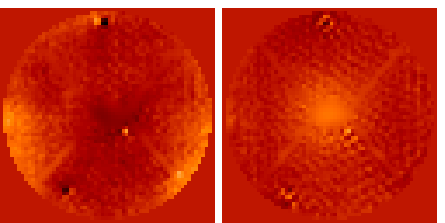}\\
\includegraphics[width = 0.95\linewidth]{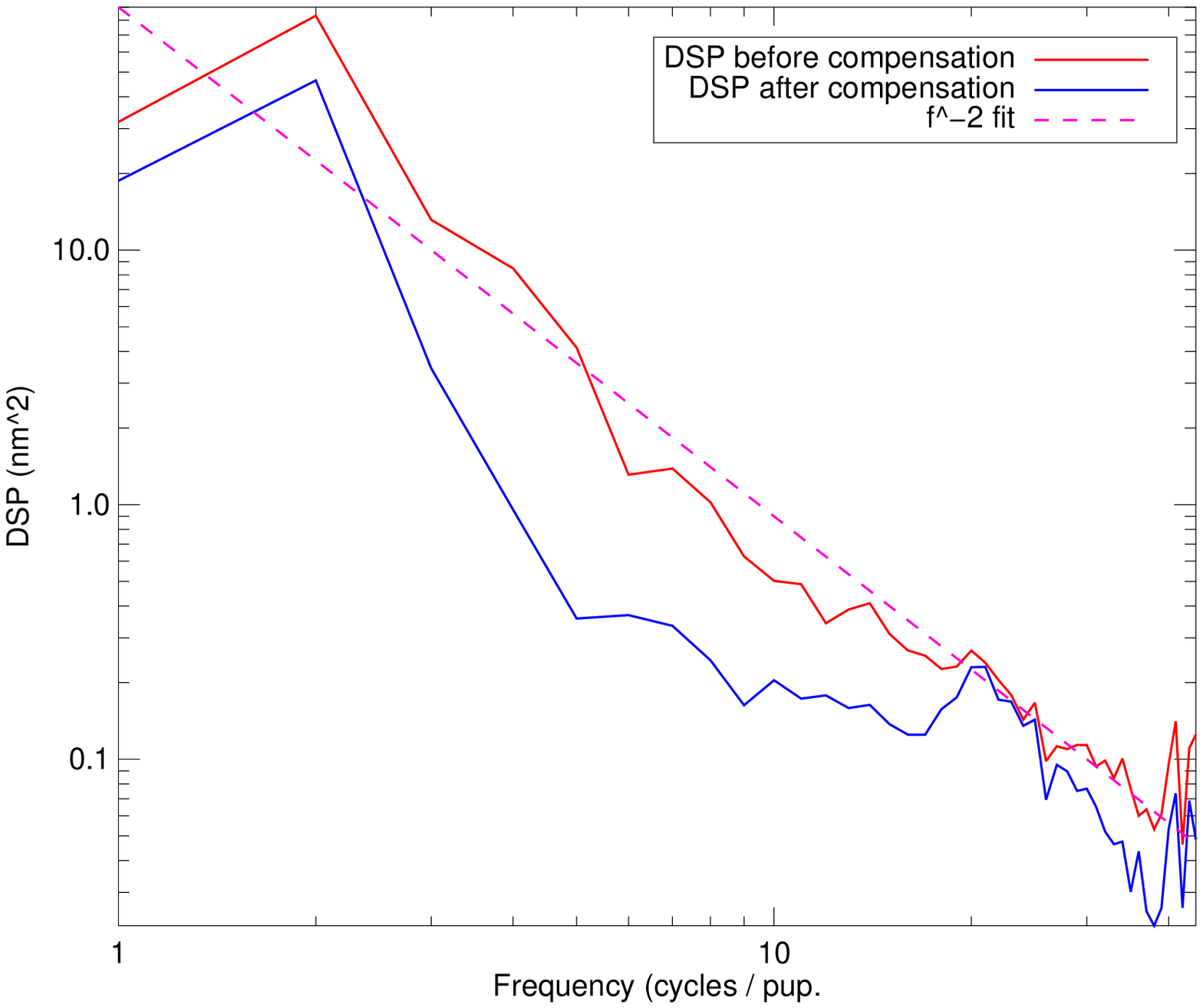}\\
\end{tabular}
\caption{Residual quasi-static aberrations upstream of the coronagraph on
  SPHERE. Up: aberrations upstream of the coronagraph estimated by COFFEE
  before (left, WFE$= 106$ nm RMS) and after compensation (right, WFE$=76$ nm
  RMS, same dynamic for both images). Bottom: PSD profiles computed for the
  estimated aberration before (solid red line) and after (solid blue line)
  compensation. For comparison, a $1/f^2$ law is plotted (dashed magenta
  line).}
\label{dsp_exp}
\end{figure}

The aberrations estimated by COFFEE before and after compensation during the
PCL process (Section \ref{sect_comp_exp}) have now been analyzed. In Figure
\ref{dsp_exp}, which presents both the estimated aberrations and their PSD,
the impact of the PCL process performed in Section \ref{sect_comp_exp} can
clearly be seen in the estimated aberration map. The aberration level indeed
decreases after compensation, which is quantified by the decreasing aberration
WFE. Such behavior is confirmed by the PSD (figure \ref{dsp_exp}, bottom) of
these aberrations: thanks to the PCL, all frequencies controlled by SAXO, from
$0$ to $20$ cycles per pupil (which is SAXO's cut-off frequency) decrease.
Again, such a result demonstrates COFFEE's usefulness for SPHERE compared to
the classical phase diversity, whose own cut-off frequency is eight cycles per
pupil.

In Figure \ref{dsp_exp}, the PSD of the aberrations estimated before
compensation show the same $1/\nu^2$ behavior as the one found by
\cite{Dohlen-p-11}. This result demonstrates that COFFEE can be used as a
simple and powerful tool for characterizing SPHERE's residual aberrations. Indeed,
this wavefront sensing method requires only two images to perform the
aberration estimation and can be used with the same settings as the ones
required by the scientific observation, in particular the coronagraph (which
has to be removed for a measurement with a conventional wavefront sensor).

\section{Conclusion}
\label{sect_ccl}

In this paper, we have used COFFEE, the coronagraphic phase diversity method,
to estimate and compensate for SPHERE's quasi-static aberrations, leading to a
contrast optimization on the IRDIS detector. In Section \ref{sect_coffee}, the
application of COFFEE to SPHERE was presented and used in Section
\ref{sect_est} to demonstrate the ability of COFFEE to estimate high-frequency
aberrations. We introduced an influence function pattern with SAXO's HODM and
then estimated it with a nanometric precision using COFFEE from coronagraphic
images recorded from the IRDIS detector. In Section \ref{sect_comp}, COFFEE
was used to compensate for SPHERE's own quasi-static aberrations. We developed
a refined compensation process to modify SAXO's references slopes using
COFFEE's estimation, which allowed us to optimize the contrast up to a factor
$4.7$ on the IRDIS detector. 

We compared the measured contrast gain with simulations that closely mimic the
SPHERE instrument and demonstrated that the compensation was limited by SAXO's
performance. In particular, we have shown that \review{the control basis
  truncation, performed to take dead actuators and to improve SAXO's
  robustness,} was responsible for limiting of the achievable contrast.
Finally, the residual aberrations estimated by COFFEE were analyzed,
demonstrating that this method could also be used as a simple and powerful
tool for measuring and characterizing SPHERE's residual aberrations.

Several perspectives are currently considered to optimize the control of
SPHERE's quasi-static aberrations. First, the compensation stage used in the
PCL process should be improved to allow an \review{even greater} contrast
gain. Two solutions are currently considered. The first one consists in a
modified control loop that would use a control matrix where less KL modes
would be filtered out. Such a control matrix, which would be used only for the
system calibration (and not during the scientific observation) would allow one
to compensate for the aberrations that corresponds to the KL modes that were
filtered out in this paper, leading to an improved contrast gain. Still, such
a compensation would be limited by dead actuators, which limits the achievable
contrast \review{expecially far from the optical axis}, as demonstrated in
this paper. Improved performance can be achieved by using dark hole methods,
such as the one proposed by \cite{Paul-a-13b}, which consists in minimizing
the energy in a selected area on the detector. Notably, it should be possible
to perform this minimization over the $999$ KL modes controlled by SAXO, which
would lead to a far better contrast on the detector than the one achievable
using conventional phase conjugation. Further perspectives include deriving
the regularization metrics from the analysis of the intensity distribution in
the coronagraphic image. Besides, to optimize the dark hole--based
compensation, we are currently working on an extension of COFFEE able to
estimate amplitude aberrations. \review{Preliminary simulations suggest that
  in order to estimate the amplitude aberration along with phase aberrations,
  COFFEE will require a third coronagraphic image, created by introducing
  another diversity phase $\bphi_{\inddiv_2}$ upstream of the coronagraph.}

\begin{acknowledgements}
  The authors would like to thank the R\'egion Provence-Alpes-C\^ote d'Azur
  for partial financial support of B. Paul's scholarship. This work was partly
  funded by the European Commission under FP7 Grant Agreement No. 312430
  Optical Infrared Coordination Network for Astronomy, and by the Office
  National d’\'Etudes et de Recherches A\'erospatiales (ONERA) in the
  framework of the NAIADE Research Project. SPHERE is an instrument designed
  and built by a consortium consisting of IPAG, MPIA, LAM, LESIA, Laboratoire
  Fizeau, INAF, Observatoire de Geneve, ETH, NOVA, ONERA and ASTRON in
  collaboration with ESO.
\end{acknowledgements}

\end{document}